\documentstyle[epsfig]{aipproc}
\begin{document}
\title{Fishing for Narrow Dibaryons in $pd\to pX$ Reaction }

\author{L.V. Fil'kov$^a$, V.L. Kashevarov$^a$, E.S. Konobeevski$^b$, M.V. 
Mordovskoy$^b$, 
S.I. Potashev$^b$, V.A. Simonov$^b$, V.M. Skorkin$^b$, and S.V. Zuev$^b$}
\address{$^a$Lebedev Physical Institute, 117924 Moscow, Russia\\
$^b$Institute for Nuclear Research, 117312 Moscow, Russia}
\maketitle
\begin{abstract}
An analysis of new experimental data, obtained at Linear 
Accelerator of INR, is carried out with the aim of searching for
supernarrow dibaryons in the reactions $pd\to p+X$ and $pd\to p+pX_1$. 
Dibaryons with
masses 1904$\pm 2$, 1926$\pm 2$, and 1942$\pm 2$ MeV have been observed 
in missing mass $M_{X}$ spectra.
In missing mass $M_{X_1}$ spectra, the resonancelike states 
$X_1=\gamma+n$ at $M_{X_1}=966\pm 2$, 986$\pm 2$, and 1003$\pm 2$ MeV 
have been found. The analysis of the data obtained leads to the 
conclusion that the observed dibaryons are supernarrow dibaryons, the
decay of which into two nucleons is forbidden by the Pauli exclusion 
principle.
\end{abstract}

In Ref.\cite{ksf,yad,prc} the study of the reaction $pd\to pX$ was
performed with the aim of searching for supernarrow dibaryons (SND), the
decay of which into two nucleons is forbidden by the Pauli exclusion
principle \cite{mul,fil1,fil2}. Such dibaryons with the mass
$M<2m_N+m_{\pi}$ can decay into two nucleons, mainly emitting 
a photon. The experiment was 
carried out at 305 MeV using
the two-arm spectrometer TAMS. As was shown in Ref. \cite{yad,prc},
the nucleons and the deuteron from the decay of SND into $\gamma NN$
and $\gamma d$ have to be emitted in a narrow angle cone with respect to
the direction of motion of the dibaryon. On the other hand, if a dibaryon
decays mainly into two nucleons, then the expected  angular cone of
emitted nucleons must be more than $50^{\circ}$. Therefore, a detection
of the scattered proton in coincidence with the proton (or the deuteron)
from the decay of particle $X$ at correlated angles allowed to suppress
essentially the contribution of the background processes and to increase
the relative contribution of a possible SND production. As a result,
two narrow peaks in missing mass spectra have been observed at
$M=$1905 and 1924 MeV. The analysis of the angular distributions of the protons
from the decay of particle $X$ showed that the peak found at 1905 MeV
most likely corresponds to a SND with isotopic spin equal to 1. 
In Ref. \cite{prc} arguments were presented for the resonance at
$M=$1924 MeV is a SND, too.

In the present paper we give the results of an analysis of
new experimental data of $pd\to p+X$ 
and $pd\to p+pX_1$ reactions at 305 MeV. Experiment was performed using the 
spectrometer TAMS, the properties of which were described
elsewhere \cite{prc}. CD$_2$ and C$^{12}$ were used as targets. 
In this experiment, the scattered proton was detected in the left arm
of the spectrometer TAMS at the angle $\theta_L=70^{\circ}$. The second
charged particle (either $p$ or $d$ from the decay of $X$ state) was detected in 
the right arm by three telescopes located at $\theta_R=34^{\circ}$,
$36^{\circ}$, and $38^{\circ}$. 

As follows from the present experiment, the main contribution into 
resonances observed here 
is given by
processes where the second charged particle is a proton. 
The experimental missing mass $M_{X}$ spectra 
obtained with the CD$_2$ target are shown in Figs. 1(a-c), where (a), (b), and 
(c) correspond to a detection of the
proton from the decay of $X$ states in the right arm detector at
$\theta_R=34^{\circ}$, $36^{\circ}$, and $38^{\circ}$, respectively.
\begin{figure}[b!] 
\centerline{\epsfig{file=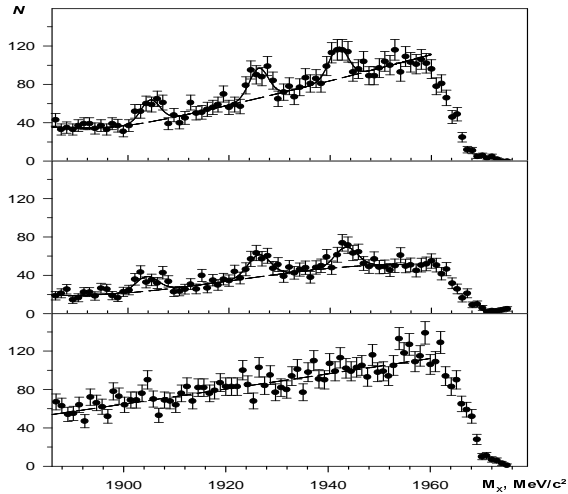,height=3.5in,width=3.5in}}
%\vspace{10pt}
\caption{The missing mass $M_{X}$ spectra for the reaction $pd\to p+X$; 
(a) -- $\theta_R=34^{\circ}$, (b) -- $\theta_R=36^{\circ}$, 
(c) -- $\theta_R=38^{\circ}$. }
\label{fig1}
\end{figure}

Three peaks at $M_{X}=1904\pm 2$, 1926$\pm 2$, and 1942$\pm 2$ MeV are 
observed in these spectra.
The first two of them confirmed the values of the dibaryon mass obtained
by us earlier \cite{ksf,yad,prc} and the resonance at 1942 MeV is a new one.  
It is expected \cite{prc,fil2} that isoscalar SNDs contribute mainly
into $\gamma d$ channel and isovector SNDs do into $\gamma NN$ one.
As the main decay of the found dibaryons is observed into $pX_1$
channel, it is possible to assume that $X_1=\gamma +n$ and all these states are 
isovector SNDs. 
The calculations for the SNDs $D(T=1,J^P=1^{\pm})$ 
showed that the biggest contribution of such dibaryons must be 
at $\theta_R=34^{\circ}$ and $36^{\circ}$. 
The contribution to spectrum at $38^{\circ}$ is expected to be several times 
smaller.
These predictions are in agreement with our experimental data.
If the observed states are usual $NN$-coupled dibaryons decaying 
mainly into two nucleons then 
their contributions to the missing mass spectra in Fig.1(a), 
1(b), and 1(c) would
be nearly the same and would not exceed a few events.
Hence, the peaks found
most likely correspond to isovector SNDs.

The summary spectrum over angles $\theta_R=34^{\circ}$ and $36^{\circ}$
is presented in Fig. 2a. This spectrum was interpolated by a
second order polynomial (for the background) plus Gaussians (for the
peaks). 
\begin{figure}[b!] 
\centerline{\epsfig{file=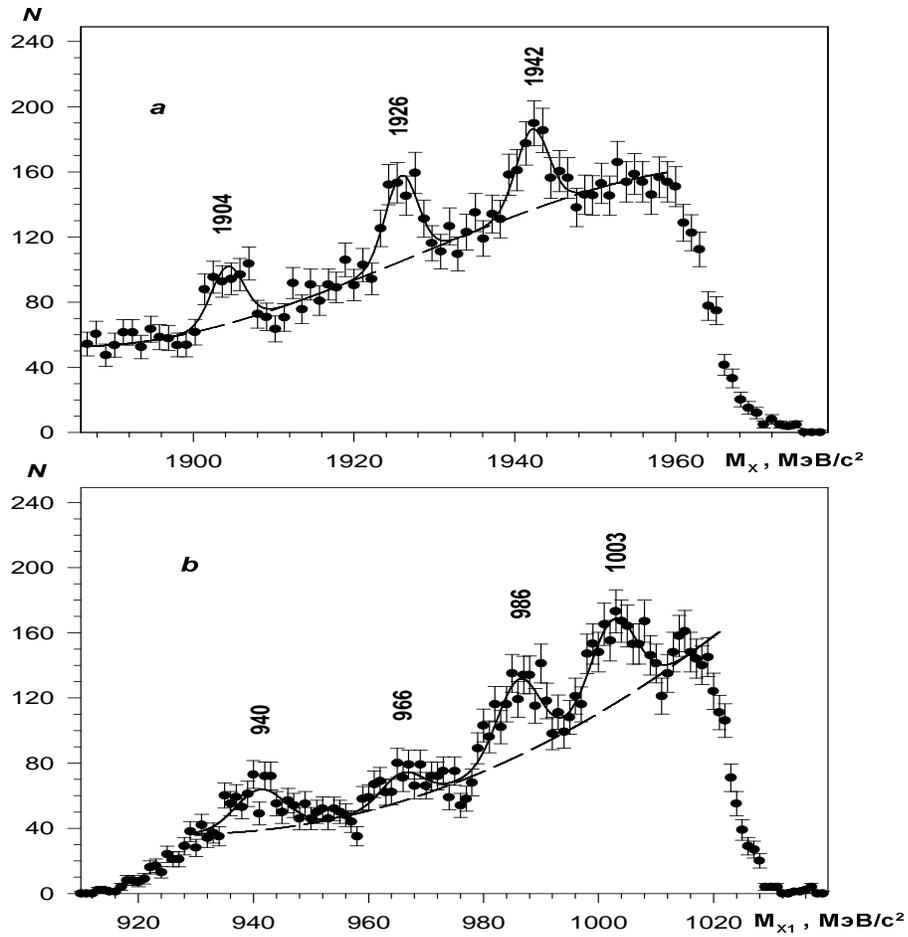,height=5.5in,width=5.5in}}
%\vspace{10pt}
\caption{ The missing mass $M_{X}$ (a) and $M_{X_1}$ (b) for the sum of angles 
of  $\theta_R=34^{\circ}$ and 
$\theta_R=36^{\circ}$.}
\label{fig2(a,b)}
\end{figure}
The numbers of standard deviations
are 6.0, 7.0, and 6.3 SD for resonances at 1904, 1926, and 1942 MeV, 
respectively. 

An additional information about the nature of the observed states is given
by
study of the missing mass $M_{X_1}$ spectra of the reaction $pd\to p+pX_1$.
If the state found is a dibaryon decaying mainly into two nucleons then
$X_1$ is a neutron and the mass $M_{X_1}$ is equal to the neutron mass
$m_n$. If the value of $M_{X_1}$, obtained from the experiment, differs
essentially from $m_n$ then $X_1=\gamma+n$ and we have the additional 
indication that the observed dibaryon is SND. 

The simulation of mass spectra for the reaction $pd\to p+pX_1$, where
$pX_1$ are decay products of SNDs with masses 1904, 1926, and 1942 MeV, gave 
peaks
at $M_{X_1}$=965, 987,and 1003 MeV, respectively.
Fig. 2b demonstrates the missing mass $M_{X_1}$ spectrum obtained from the
experiment for the sum of the angles $\theta_R=34^{\circ}$ and $36^{\circ}$.
As is seen from this figure, besides the peak at neutron mass, 
which caused by the process $pd\to p+pn$,
resonance-like behavior of the spectrum is observed at $966\pm 2$,
$986\pm 2$, and $1003\pm 2$ MeV. These values of $M_{X_1}$ coincide with
the ones obtained  from the simulation and differ essentially from 
the value of the neutron mass. Hence, for all states under study, 
$X_1=\gamma+n$ and the dibaryons found are really SNDs.

It should be noted that a resonance-like behavior of $X_1$ at
$M_{X_1}=1003\pm 2$ MeV corresponds to the resonance found in \cite{tat} and
attributed to an excited nucleon state $N^*$. In this work, the authors 
brought out three such states with masses 1004, 1044, and 1094 MeV. 
Taking into account the found connection between the SNDs 
and the resonance-like states $X_1$, 
it is possible to assume that the peaks, observed in \cite{tat} 
are not the excited nucleons, but
they are resonance-like states
$X_1$ caused by possible existence and decay of SNDs with the masses 1942, 
1982, and 2033 MeV, respectively. 

The following conclusion can be made. As a result of the study of the
reaction $pd\to pX$ and $pd\to p+pX_1$ three narrow peaks at 1904, 1926,
and 1942 MeV have been observed in the missing mass $M_{X}$ spectra.
The analysis of the angular distributions of the protons from decay of
$X$ states showed that the peaks found can be explained as a 
manifestation of the SNDs, the decay of which into two nucleons is 
forbidden by the Pauli exclusion principle.
The observation of the resonance-like structures in the missing mass
$M_{X_1}$  spectra at 966, 986, and 1003 MeV is an additional confirmation
that the dibaryons found are really SNDs.


\begin{references}
\bibitem{ksf} Kashevarov V.L., Konobeevski E.S., Mordovskoy M.B.,
Potashev S.I., Skorkin V.M., and Fil'kov L.V., {\it Bulletin of 
Lebedev Phys. Inst.} No {\bf 11}, 36 (1998).
\bibitem{yad} Fil'kov L.V., Kashevarov V.L., Konobeevski E.S.,
Mordovskoy M.V., Potashev S.I., and Skorkin V.M.,
{\it Phys. Atom. Nucl.} {\bf 62}, 2021, (1999).
\bibitem{prc} Fil'kov L.V., Kashevarov V.L., Konobeevski E.S.,
Mordovskoy M.V., Potashev S.I., and Skorkin V.M., {\it Phys. Rev. C} {\bf 61},
044004, (2000).
\bibitem{mul} Mulders P.J., Aerts A.T., and de Swart J.J., {\it Phys. Rev. D}
21, 2653, (1980).
\bibitem{fil1} Fil'kov L.V., {\it Sov. J. Nucl. Phys.} {\bf 47}, 437, (1988).
\bibitem{fil2} Akhmedov D.M. and Fil'kov L.V., {\it Nucl. Phys.} {\bf A544},
692, (1992).
\bibitem{tat} Tatischeff B. {\em et al.} {\it Phys. Rev. Lett.} {\bf 79}, 601,
(1997).  
\end{references}
\end{document}